\documentclass{IEEEtran}\usepackage[font=footnotesize]{caption}
\IEEEoverridecommandlockouts
\usepackage{cite}
\usepackage{amsmath,amssymb,amsfonts,upgreek}
\usepackage{algorithmic}
\usepackage{graphicx}
\usepackage{textcomp}
\usepackage{xcolor}
\usepackage{balance}
\usepackage{subcaption} 
\usepackage{float}
\DeclareMathOperator*{\argminA}{arg\,min}

\def\BibTeX{{\rm B\kern-.05em{\sc i\kern-.025em b}\kern-.08em
    T\kern-.1667em\lower.7ex\hbox{E}\kern-.125emX}}
\begin{document}

\title{Complex-Valued Neural Networks for \\ Ultra-Reliable Massive MIMO\\}

\author{\IEEEauthorblockN{Pedro B. Valadares, Jonathan A. Soares, Kayol S. Mayer, and Dalton S. Arantes}
\IEEEauthorblockA{\textit{Department of Communications, School of Electrical and Computer Engineering} \\
\textit{Universidade Estadual de Campinas -- UNICAMP}, Campinas, SP, Brazil\\
p204483@dac.unicamp.br, jonathan@decom.fee.unicamp.br, kayol@unicamp.br, dalton@unicamp.br}
}

\maketitle

\begin{abstract}
In the evolving landscape of 5G and 6G networks, the demands extend beyond high data rates, ultra-low latency, and extensive coverage, increasingly emphasizing the need for reliability. This paper proposes an ultra-reliable multiple-input multiple-output (MIMO) scheme utilizing quasi-orthogonal space-time block coding (QOSTBC) combined with singular value decomposition (SVD) for channel state information (CSI) correction, significantly improving performance over QOSTBC, and traditional orthogonal STBC (OSTBC) when analyzing spectral efficiency. Although QOSTBC enhances spectral efficiency, it also increases computational complexity at the maximum likelihood (ML) decoder. To address this, a neural network-based decoding scheme using phase-transmittance radial basis function (PT-RBF) architecture is also introduced to manage QOSTBC’s complexity. Simulation results demonstrate improved system robustness and performance, making this approach a potential candidate for ultra-reliable communication in next-generation networks.
\end{abstract}


\section{Introduction}

The exponential growth in global data consumption, driven by technologies such as 5G, cloud computing, and the Internet of Things (IoT), creates increasing demand for higher spectral efficiency, robust telecom infrastructure, and enhanced reliability. While many 5G and upcoming 6G network applications focus on delivering higher data rates, not all use cases prioritize speed. For instance, autonomous vehicles rely heavily on the reliability of data transmission to ensure safe and efficient operation, making reliability a critical factor. However, these systems often face performance challenges due to multipath propagation and additive white Gaussian noise~(AWGN)~\cite{chen2020_6G_Autonomous,wang2021_antennas_V2X,samimi2017_ultra_reliable_MIMO}.

To address these challenges, implementing transmit diversity through space-time block coding~(STBC) has shown promise, as it improves signal robustness against fading and mitigates issues related to multipath propagation. STBC achieves it by transmitting the encoded data across multiple antennas, allowing the system to take advantage of spatial diversity and significantly boost communication reliability and signal quality. This method not only compensates for channel-related impairments but also optimizes the system’s use of available bandwidth and power, enhancing the overall performance of modern telecom networks~\cite{tang2019svd}.

STBC is commonly used in various communication systems to enhance performance. However, one of the key challenges in deploying STBC lies in the complexity of creating orthogonal coding schemes~(OSTBC) when several transmitting antennas are involved. Although OSTBC delivers strong diversity gains, its practical application becomes more difficult with an increasing number of antennas, as achieving orthogonality becomes more complex. QOSTBC provides an alternative with better spectral efficiency due to its higher code rate and also enables the implementation of a higher number of antennas than OSTBC. On the other hand, it sacrifices some of the diversity gains seen with OSTBC. Additionally, QOSTBC requires more complex computations during decoding, as it involves processing multiple transmitted symbols using maximum likelihood decoders~\cite{STBC_MassiveMIMO_16}.

In order to circumvent the prohibitive computational complexity of QOSTBC ML decoding, Wang et al.~\cite{Wang2022} and Soares et al.~\cite{soares2021} independently proposed the use of neural networks for joint channel estimation and decoding. While the real-valued neural networks~(RVNNs) of Wang et al.~\cite{Wang2022} were limited to systems with $4\times4$ antennas and 4-QAM (quadrature amplitude modulation), the complex-valued neural network~(CVNN) of Soares et al.~\cite{soares2021} could handle $32\times32$ antennas and 64-QAM, marking the first work to address quasi-orthogonal STBC~(QOSTBC) systems with more than $4\times4$ antennas and higher-order QAM modulations.

This paper proposes a novel approach that integrates an SVD-based CSI correction with QOSTBC to improve robustness and performance in MIMO systems and enable these systems to work in massive antenna arrangements, breaking the limit imposed by traditional OSTBC. To address the computational challenge related to linear decoding with QOSTBC, we employ phase transmittance radial basis function~(PTRBF) neural networks at the receiver. This method effectively mitigates the decoding complexity, offering a robust solution to the challenges of QOSTBC. The results validate the success of this innovative approach in improving system reliability and decoding efficiency, demonstrating its potential as a practical solution for next-generation communication systems.

The remainder of this paper is organized as follows: Section~\ref{sec:system_model} provides a brief review of STBC systems. The proposed parallel channel estimation and decoding strategy for MIMO OFDM systems are detailed in Section~\ref{sec:proposed_QOSTBC_SVD} and the improvement with a CVNN channel estimation and decoding is provided in Section~\ref{sec:proposed_approach}. Conclusions are discussed in Section~\ref{sec:conclusions}.

\section{MIMO-STBC System Model}
\label{sec:system_model}

This paper considers the MIMO space-diversity scheme based on space-time block coding (STBC). STBC enhances communication reliability by transmitting multiple copies of the signal through multiple antennas. As the number of antennas increases, the likelihood that all signal copies will be affected by deep fading is significantly reduced\cite{alam2021_performance_STBC_MIMO}.

\subsection{Space-Time Block Coding}
\label{sub:STBC}

Space-time block codes (STBCs)~\cite{Jankiraman2004,Tarokh1999,Li2021}, can achieve the full-transmit diversity of $N_{tx} N_{rx}$, where $N_{tx}$ and $N_{rx}$ denote the numbers of transmitting antennas and receiving antennas, respectively. This optimal diversity performance can be achieved by employing the maximum likelihood~(ML) decoding at the receiver~\cite{Jankiraman2004}. 

The idea is to transmit $N_{tx}$ orthogonal streams, which ensures that the receiver antennas receive $N_{tx}$ orthogonal streams. This special class of space-time block codes is known as orthogonal STBC (OSTBC)~\cite{soares2021,Tarokh1999,Hu2020}. Besides diversity gain, the OSTBC leads to a secondary linear coding gain $G_c = 10\log\left(R\right)$ at the receiver due to the coherent detection of multiple signal copies over time, and an array gain $G_a = 10\log\left(N_{rx}\right)$ due to the coherent combination of multiple received signals over the receiving antennas~\cite{soares2021}.

One of the disadvantages of OSTBC is the code rate. Let $N_{tp}$ represent the number of time samples to convey one block of coded symbols, and $N_s$ represent the number of symbols transmitted per block. The space-time block code rate is defined as the ratio between the number of symbols that the encoder receives at its input and the number of space-time coded symbols transmitted from each antenna, given by $R = N_s/N_{tp}$. For example, an OSTBC coding matrix for $N_{tx}=4$ implies a code rate $R=1/2$, limiting spectral efficiency.

\subsection{Quasi-Orthogonal Space-Time Block Coding}

In order to increase the spectral efficiency in orthogonal codes, Jafarkhani~\cite{Jafarkhani898239} proposed quasi-orthogonal STBC (QOSTBC) of rate one, relaxing the requirement of orthogonality. However, when compared with orthogonal codes, the diversity gain is reduced by a factor of two. In contrast to orthogonally designed codes that process one symbol at a time on the decoder, quasi-orthogonal codes process pairs of transmitted symbols, which exponentially increases the computational complexity of decoding~\cite{soares2021}. Jafarkhani~\cite{Jafarkhani898239} proposed a coding matrix of rate one for $N_{tx}=4$, given by

\begin{equation}
    \label{eq:QOSTBC44}
    \mathbf{QOSTBC_{4,4}}=\begin{bmatrix}
   s[1] & s[2] & s[3] & s[4]\\ 
   -s[2]^* & s[1]^* & -s[4]^* & s[3]^*\\
   -s[3]^* & -s[4]^* & s[1]^* & s[2]^*\\
   s[4] & -s[3] & -s[2] & s[1]\\
    \end{bmatrix}.
\end{equation}

In the literature, related approaches with a maximum of $N_{tx}=6$ antennas have been proposed for quasi-orthogonal codes~\cite{TBH876470,Weifeng1188366,SindhuHameed7383923}. In~\cite{Weifeng1188366}, the authors developed an architecture similar to~\cite{Jafarkhani898239}. However, it presents full diversity at the cost of more processing and is limited to $N_{tx}=4$ antennas. Similarly, by increasing the decoding processing, Sindhu and Hameed~\cite{SindhuHameed7383923} proposed two quasi-orthogonal schemes with $N_{tx}=5$ and $6$ antennas~\cite{soares2021}.

\subsection{Generalized Quasi-Orthogonal Coding Scheme}

In this paper, we use the generalized recursive method proposed in~\cite{soares2021} for generating QOSTBC coding schemes:
\begin{equation}
    \mathbf{S}_{N_s}^{N_{tx}} = 
    \begin{bmatrix}
    \mathbf{S}_{N_s-N_{tx}/2}^{N_{tx}/2} & \mathbf{S}_{N_s}^{N_{tx}/2}\\ 
    -[{\mathbf{S}_{N_s}^{N_{tx}/2}}]^* & [{\mathbf{S}_{N_s-N_{tx}/2}^{N_{tx}/2}}]^*
    \end{bmatrix},
    \label{eq:propCode}
\end{equation}
in which $N_{tx} = 2^n$, $\forall \: n \in \mathbb{N}^+$, is the number of transmitting antennas and $N_s$ is the number of encoded symbols. In this encoding approach, $N_s\triangleq N_{tx}$ and the code rate is $R=N_{tx}/N_s = 1$~\cite{soares2021}. The recurrence is employed until ${\mathbf{S}_{n}^{1}}=s[n], \, \forall n \in [1,\,2,\cdots,N_s]$ in~\eqref{eq:propCode}.

With four transmitting antennas,~\eqref{eq:propCode} results in:

\begin{equation}
    \label{eq:S4}
    \mathbf{S}_{4}^{4} =
    \begin{bmatrix}
        s[1] & s[2] & s[3] & s[4] \\ 
        -s[2]^* & s[1]^* & -s[4]^* & s[3]^* \\ 
        -s[3]^* & -s[4]^* & s[1]^* & s[2]^* \\ 
        s[4] & -s[3] & -s[2] & s[1]
    \end{bmatrix}.
\end{equation}

Note that \eqref{eq:S4} is equal to the coding scheme proposed by~\cite{Jafarkhani898239} with four antennas, as in~\eqref{eq:QOSTBC44}. However, in contrast to the work of~\cite{Jafarkhani898239}, this scheme can generate coding matrices for any $N_{tx} = 2^n$, $\forall \: n \in \mathbb{N}^+$, and $N_s\triangleq N_{tx}$~\cite{soares2021}. For the case of $n=1$, \eqref{eq:propCode} is equal to the Alamouti coding, the full-rate full-diversity complex-valued space-time block code proposed in~\cite{Alamouti730453}.

\subsection{System Model}

\begin{figure}[htbp]
\begin{center}
    \includegraphics[width=\columnwidth]{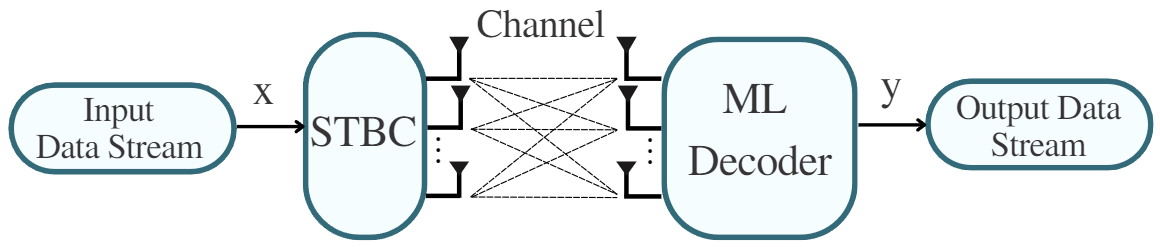}
\caption{Space-time block coding configuration for multiple-input multiple-output (MIMO) system with ML decoder.}
    \label{fig:MIMO_Model}
\end{center}
\end{figure}

Fig.~\ref{fig:MIMO_Model} illustrates a basic diagram of the MIMO-STBC scheme under consideration. In the input data stream block, a sequence of bits is mapped onto an $M$-QAM constellation denoted by $\mathfrak{C}$. The transmitting space-time block coder (STBC) then processes the stream of QAM symbols, utilizing a code matrix to generate a transmission matrix $\mathbf{X}\in \mathbb{C}^{N_{tx}\times P}$, where $N_{tx}$ denotes the number of transmitting antennas and $P$ represents the code matrix length.

Considering a flat fading, time-invariant channel in a MIMO-STBC system, the received symbols can be expressed as:
\begin{equation}
    \mathbf{Y} = \mathbf{H}^T\mathbf{X}+\mathbf{Z}, \quad \mathbf{Y} \in \mathbb{C}^{N_{rx}\times P},
\end{equation}
in which $\mathbf{Z}\in \mathbb{C}^{N_{rx}\times P}$ represents the additive white Gaussian noise (AWGN) at the $N_{rx}$ receiving antennas, and $\mathbf{H}\in \mathbb{C}^{N_{tx}\times N_{rx}}$ is the channel matrix.

At the receiver, the ML decoder, using the estimated channel matrix $\widehat{\mathbf{H}}$, decodes the STBC matrices into QAM symbols, which are subsequently demapped into bits at the output data stream block.

However, for QOSTBC, the decoding complexity increases due to the need to process pairs of transmitted symbols. To address this challenge, we propose employing non-linear decoding methods at the receiver, such as complex-valued neural networks (e.g., phase-transmittance radial basis function (PT-RBF) neural networks~\cite{soares2021}), which can effectively handle the increased complexity.


\section{Proposed QOSTBC SVD-Based Model}
\label{sec:proposed_QOSTBC_SVD}

In this section, we propose an enhanced QOSTBC MIMO system utilizing singular value decomposition (SVD)-based precoding to improve system performance and robustness.The core concept involves precoding the QOSTBC-encoded signal using the singular vectors of the channel matrix, thereby aligning the transmission with the most favorable channel modes, similar to beamforming.

\subsection{Channel Decomposition and Precoding}

When channel state information (CSI) is available at the transmitter, the capacity of a MIMO system can be enhanced using singular value decomposition (SVD)~\cite{Raleigh_TC_1998}. The SVD of the channel matrix transpose $\mathbf{H}^T$ is given by:

\begin{equation}
    \mathbf{H}^T = \mathbf{U} \mathbf{\Sigma} \mathbf{V}^H,
\end{equation}
in which $\mathbf{U} \in \mathbb{C}^{N_{rx} \times N_{rx}}$ and $\mathbf{V} \in \mathbb{C}^{N_{tx} \times N_{tx}}$ are unitary matrices containing the left and right singular vectors of $\mathbf{H}^T$, respectively, and $\mathbf{\Sigma} \in \mathbb{C}^{N_{rx} \times N_{tx}}$ is a diagonal matrix containing the singular values $\sigma_i$ of $\mathbf{H}^T$.

By employing SVD-based precoding, the transmitter can align the transmitted signals with the dominant singular vectors of the channel, effectively transmitting in the directions/components of the highest energy. This process enhances the received signal strength and improves overall system performance.

\subsection{Precoding of QOSTBC Coded Signals}

Let $\mathbf{S}$ denote the QOSTBC coded block generated as per the generalized scheme in~\eqref{eq:propCode}. To perform precoding, we multiply the coded block $\mathbf{S}$ by the precoding matrix $\mathbf{V}$ obtained from the SVD of the channel's CSI:

\begin{equation}
    \mathbf{X} = \mathbf{V} \mathbf{S},
\end{equation}
where $\mathbf{X} \in \mathbb{C}^{N_{tx} \times P}$ is the precoded transmission matrix.

The received signal at the receiver becomes:

\begin{equation}
    \mathbf{Y} = \mathbf{H}^T \mathbf{X} + \mathbf{Z} = (\mathbf{H}^T \mathbf{V}) \mathbf{S} + \mathbf{Z}.
\end{equation}

Since $\mathbf{H}^T = \mathbf{U} \mathbf{\Sigma} \mathbf{V}^H$, we have:
\[
\mathbf{H}^T \mathbf{V} = (\mathbf{U} \mathbf{\Sigma} \mathbf{V}^H) \mathbf{V}.
\]

Therefore:
\[
\mathbf{Y} = \mathbf{U} \mathbf{\Sigma} \mathbf{S} + \mathbf{Z}.
\]

Given that $\mathbf{V}$ and  $\mathbf{U}$ are unitary, i.e., $\mathbf{V}^H \mathbf{V} = \mathbf{I}$ and $\mathbf{U}^H \mathbf{U} = \mathbf{I}$, we can simplify the expression by performing a left multiplication by left singular value hermitian $\mathbf{U}^H$ before decoding, and the effective channel becomes diagonalized, allowing for simplified decoding at the receiver, as follows: 
\begin{equation}
    \mathbf{\hat{S}} = \argminA_{\mathbf{\tilde{S}}} \left\| \mathbf{Y} - \mathbf{\Sigma} \, \mathbf{\tilde{S}} \right\|^2,
\end{equation}
in which $\mathbf{\tilde{S}} \in \mathbb{C}^{N_{tx}\times P}$ is given from the set of QOSTBC coded blocks.

\subsection{Performance Analysis}

\begin{figure}[!ht]
\centering
\includegraphics[width=\linewidth]{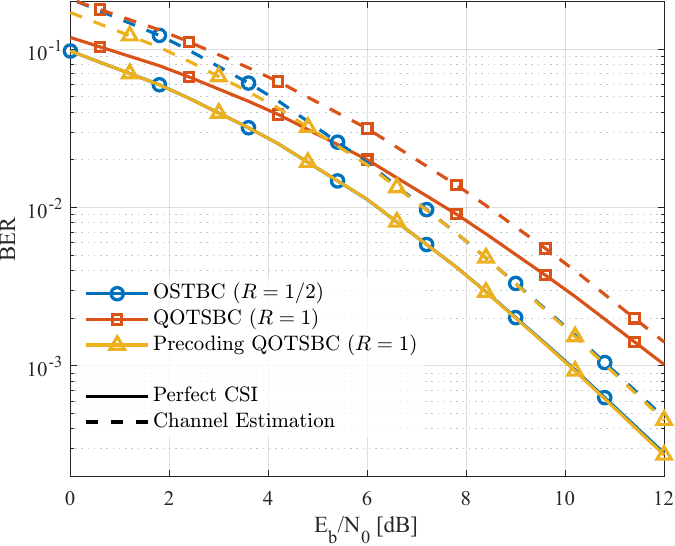}
\caption{BER performance comparison for 4-QAM MIMO systems with $N_{tx}=4$ and $N_{rx}=1$. Solid lines indicate perfect CSI; dashed lines indicate MMSE channel estimation.}
\label{fig:linear}
\end{figure}

Fig.~\ref{fig:linear} illustrates the bit error rate (BER) performance of the proposed precoded QOSTBC system compared to traditional OSTBC and unprecoded QOSTBC systems. The solid lines represent the performance with maximum likelihood (ML) decoding and perfect CSI, while the dashed lines represent the performance considering minimum mean square error (MMSE) channel estimation.

In Fig.~\ref{fig:linear}, we observe that OSTBC (blue line) achieves the best performance since it attains full diversity with a diversity order of $D_o = 4$. In contrast, QOSTBC (red line) shows lower performance due to achieving only half of the available diversity (diversity order $D_o = 2$); the relaxation of orthogonality in QOSTBC reduces the diversity gain. However, when applying SVD-based precoding to QOSTBC (yellow line), there is a significant performance improvement compared to the unprecoded QOSTBC. The precoding aligns the transmission with the strongest channel modes, effectively enhancing the diversity gain and achieving performance similar to OSTBC.

Although the precoded QOSTBC does not mathematically guarantee full diversity, the empirical results indicate that it closely approaches the performance of OSTBC by leveraging the channel's dominant singular vectors. This demonstrates that SVD-based precoding effectively mitigates the diversity loss inherent in QOSTBC, making it a promising technique for ultra-reliable communications.

The proposed SVD-based precoding for QOSTBC combines the spectral efficiency benefits of QOSTBC with the enhanced performance provided by precoding. By utilizing the channel's singular vectors, the system effectively performs a form of beamforming, directing energy along the most favorable paths and improving robustness against channel impairments.

Moreover, the application of precoding does not significantly increase the computational complexity at the transmitter, as the SVD can be computed based on the available CSI, which is typically required for other transmission strategies as well. At the receiver, the decoding complexity remains manageable, especially when combined with advanced decoding methods such as neural network-based decoders.

\section{Proposed Ultra-Reliable MIMO System with PT-RBF Neural Networks}
\label{sec:proposed_approach}

While the previous results presented in Section~\ref{sec:proposed_QOSTBC_SVD} are promising, there are well-known issues with QOSTBC systems, as pointed out by Soares et al.~\cite{soares2021}. Specifically, QOSTBC can only work effectively with phase-shift keying (PSK) modulations and the computational complexity of decoding increases exponentially with the number of transmitting and receiving antennas. To address these challenges, Soares et al.~\cite{soares2021} proposed using phase-transmittance radial basis function~(PT-RBF) neural networks as decoders.

In this paper, we employ a similar approach with optimization improvements to achieve outstanding performance in ultra-reliable MIMO systems. We introduce a neural network-based decoder that jointly estimates and decodes the received signals, mitigating the limitations of traditional QOSTBC decoding methods.

\subsection{Complex-Valued PT-RBF Neural Networks}
\label{sec:cvnns}

Following the notation used in~\cite{Mayer_LWC2022}, the PT-RBF neural network is defined with $L$ hidden layers (excluding the input layer), where the superscript $l \in [0,\,1,\,\cdots,\, L]$ denotes the layer index and $l=0$ is the input layer. The $l$-th layer (excluding the input layer $l=0$) comprises $I^{\{l\}}$ neurons, $O^{\{l\}}$ outputs, and has a matrix of synaptic weights $\mathbf{W}^{\{l\}}\in\mathbb{C}^{O^{\{l\}}\times I^{\{l\}}}$, a bias vector $\mathbf{b}^{\{l\}}\in\mathbb{C}^{O^{\{l\}}}$, a matrix of center vectors $\boldsymbol{\Gamma}^{\{l\}}\in\mathbb{C}^{I^{\{l\}}\times O^{\{l-1\}}}$, and a variance vector $\boldsymbol{\upsigma}^{\{l\}}\in\mathbb{C}^{I^{\{l\}}}$. 

Let $\mathbf{\bar{x}}\in \mathbb{C}^{N_\mathrm{inp}}$ be the normalized input vector of the PT-RBF network (${N_\mathrm{inp}}$ inputs), and $\mathbf{y}^{\{L\}}\in \mathbb{C}^{N_\mathrm{out}}$ be the output vector (${N_\mathrm{out}}$ outputs). The output vector $\mathbf{y}^{\{l\}} \in \mathbb{C}^{O^{\{l\}}}$ of the $l$-th hidden layer is given by

\begin{equation}
\label{eq:deepPTRBFNN_output}
    \mathbf{y}^{\{l\}}=\mathbf{W}^{\{l\}}\boldsymbol{\upphi}^{\{l\}}+\mathbf{b}^{\{l\}}, 
\end{equation}
in which $\boldsymbol{\upphi}^{\{l\}}\in \mathbb{C}^{I^{\{l\}}}$ is the vector of Gaussian kernels.

The $m$-th Gaussian kernel of the $l$-th hidden layer is formulated as
\begin{equation}
\label{eq:kernel}
\phi_m^{\{l\}}=\exp\left[-\Re\left(v_m^{\{l\}}\right)\right]+\jmath\exp\left[-\Im\left(v_m^{\{l\}}\right)\right],      
\end{equation}
in which $v_m^{\{l\}}$ is the $m$-th Gaussian kernel input of the $l$-th hidden layer, described as
\begin{multline}
\label{eq:kernel_argument}
v_m^{\{l\}}=\frac{\left \Vert\Re\left(\mathbf{y}^{\{l-1\}}\right)-\Re\left(\boldsymbol{\upgamma}_m^{\{l\}}\right)\right \Vert_2^2}{\Re\left(\sigma_m^{\{l\}}\right)} \\
+\jmath\frac{\left \Vert\Im\left(\mathbf{y}^{\{l-1\}}\right)-\Im\left(\boldsymbol{\upgamma}_m^{\{l\}}\right)\right \Vert_2^2}{\Im\left(\sigma_m^{\{l\}}\right)},
\end{multline}
where  $\mathbf{y}^{\{l-1\}}\in \mathbb{C}^{O^{\{l-1\}}}$ is the output vector of the $(l-1)$-th hidden layer (except for the first hidden layer where $\mathbf{y}^{\{0\}}=\mathbf{\bar{x}}$), $\boldsymbol{\upgamma}_m^{\{l\}}\in \mathbb{C}^{O^{\{l-1\}}}$ is the $m$-th vector of Gaussian centers of the $l$-th hidden layer, $\sigma_m^{\{l\}}\in\mathbb{C}$ is the respective $m$-th variance, and $\Re(\cdot)$ and $\Im(\cdot)$ return the real and imaginary components, respectively.

The PT-RBF hyperparameters were empirically optimized. For $N_{tx}=4$ and $N_{rx}=1$, the best performance was achieved with a single layer~($I^{\{1\}}=50$~neurons), while for $N_{tx}=32$ and $N_{rx}=1$, the optimal configuration used a single layer~($I^{\{1\}}=100$~neurons). In both cases, the best results were obtained with the learning rates $\eta_w^{\{1\}}=2.5\times10^{-3}$, $\eta_b^{\{1\}}=2.5\times10^{-3}$, $\eta_\gamma^{\{1\}}=1.5\times10^{-3}$, and $\eta_\sigma^{\{1\}}=2.5\times10^{-3}$ for weights, bias, center vectors, and variances, respectively.

\subsection{Performance Evaluation}

\begin{figure}[!ht]
\centering
\includegraphics[width=\linewidth]{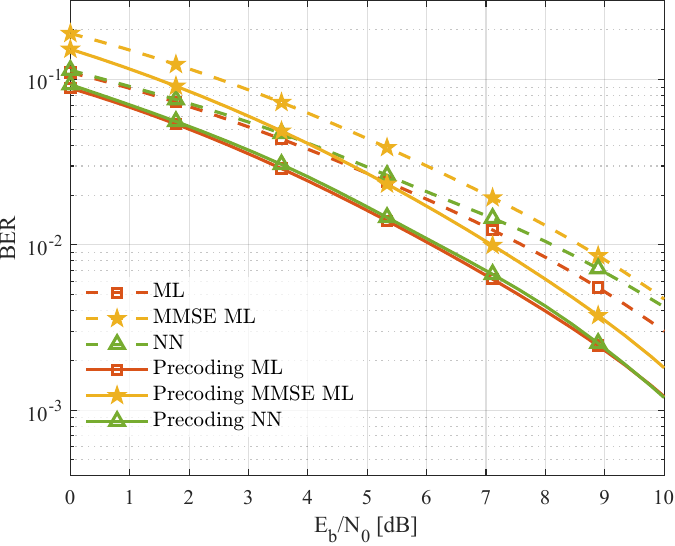}
\caption{BER performance for 4-QAM MIMO systems with $N_{tx}=4$ and $N_{rx}=1$. Solid lines indicate precoding; dashed lines indicate no precoding.}
\label{fig:NN}
\end{figure}

In Fig.~\ref{fig:NN} we show the BER performance of the proposed neural network~(NN) decoder, in a 4-QAM MIMO system with $N_{tx}=4$ and $N_{rx}=1$. This figure shows the performance of QOSTBC with ML decoding~(red line), ML decoding with MMSE channel estimation~(yellow line), and with the NN decoder~(green line). Solid lines represent systems with precoding, while dashed lines represent systems without precoding.

\begin{figure}[!ht]
\centering
\includegraphics[width=\linewidth]{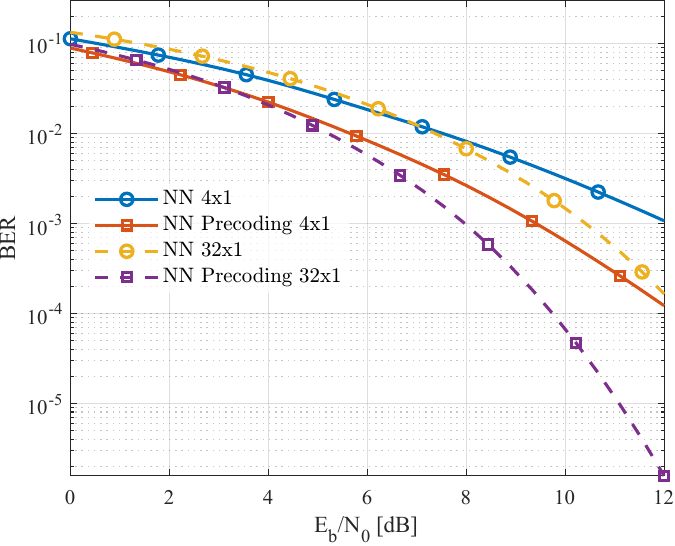}
\caption{BER performance comparison for 4-QAM massive MIMO systems with $N_{tx}=32$ and $N_{rx}=1$.}
\label{fig:massive_NN}
\end{figure}

We observe that the neural network approach closely matches the reference ML performance, with only a slight degradation. Moreover, when compared with MMSE channel estimation, the NN decoder achieves slightly better performance. This is significant because the NN decoder jointly estimates and decodes the signal, eliminating the need for separate channel estimation. As a result, it is impossible to simulate a case with perfect CSI for the NN decoder, making the comparison with the MMSE estimate scenario particularly relevant. Additional results are shown in Fig.~\ref{fig:massive_NN} for a massive MIMO system with $N_{tx}=32$ transmitting antennas, demonstrating the potential of the proposed approach to achieve the ultra-reliability required in next-generation networks, as the number of transmitting antennas increases.

\begin{figure}[b]
\centering
\includegraphics[width=\linewidth]{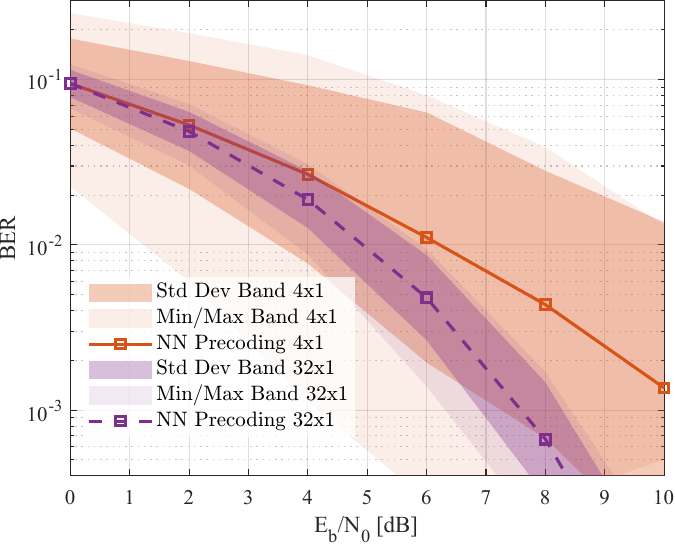}
\caption{BER performance comparison for 4-QAM massive MIMO systems with $N_{tx}=32$ and $N_{tx}=4$, standard deviation and min/max bands.}
\label{fig:NN_vs_massive_with_BER_variance_bands}
\end{figure}

Fig. \ref{fig:NN_vs_massive_with_BER_variance_bands} illustrates the variance, along with the minimum and maximum performance bands, for the NN with precoding under configurations of $4\times1$ and $32\times1$ antennas. The results demonstrate that increasing the number of antennas significantly improves the BER performance while simultaneously reducing variance. This reduction in variance ensures more consistent and predictable outcomes.

\begin{figure}[!b]
\centering
\includegraphics[width=\linewidth]{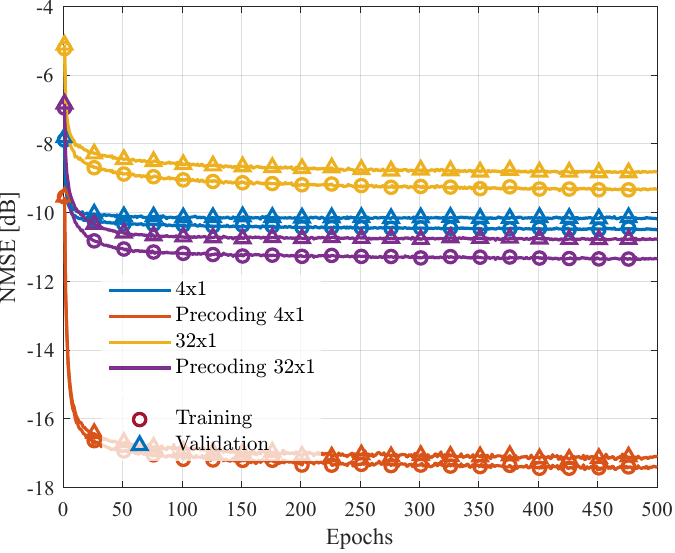}
\caption{MMSE convergence for training and validation sets under different configurations: $N_{tx}=4$ and $N_{tx}=32$ antenna arrangements, with and without SVD-based precoding.}
\label{fig:NN_vs_massive_MSE}
\end{figure}

\begin{figure}[!b]
\centering
\includegraphics[width=\linewidth]{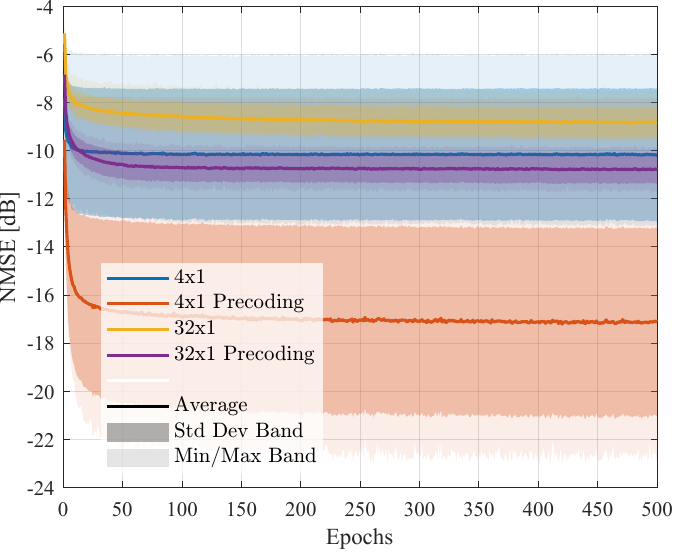}
\caption{BER performance comparison for 4-QAM massive MIMO systems with $N_{tx}=32$ and $N_{rx}=1$.}
\label{fig:NN_vs_massive_MSE_with_variance_bands}
\end{figure}

\begin{figure*}[!t]
\centering
\begin{subfigure}{0.45\textwidth}
    \centering
    \includegraphics[width=\linewidth]{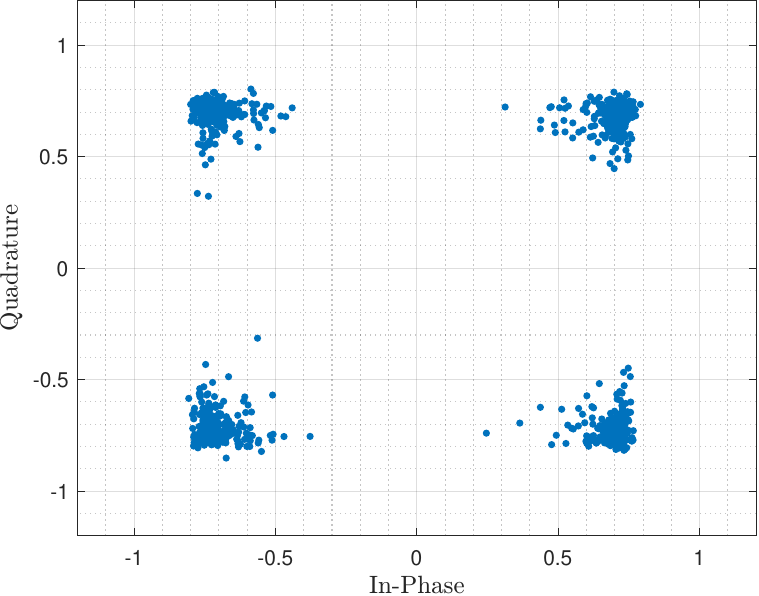}
    \caption{}
    \label{fig:Decoded_Constellation_4x1}
\end{subfigure}
\begin{subfigure}{0.45\textwidth}
    \centering
    \includegraphics[width=\linewidth]{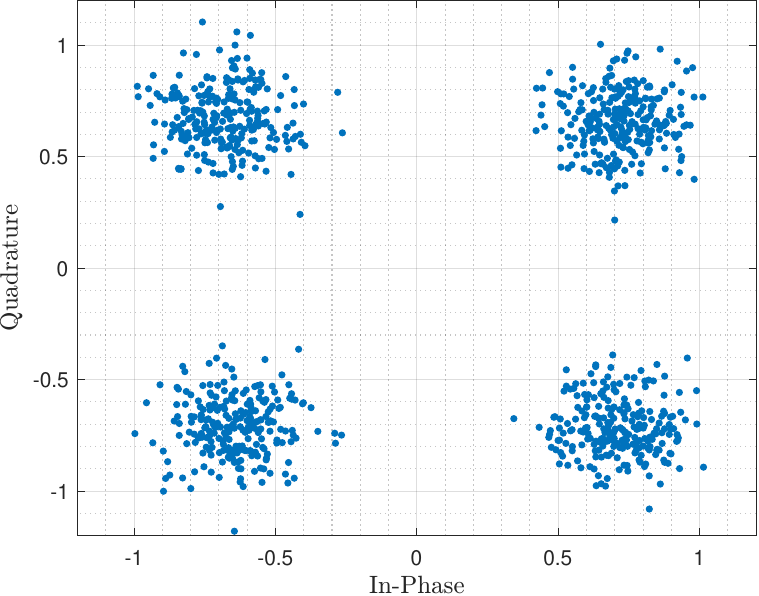}
    \caption{}
    \label{fig:Decoded_Constellation_32x1}
\end{subfigure}
\caption{Constellation Diagrams for 4-QAM MIMO PT-RBF systems: (a) $N_{tx}=4$, $N_{rx}=1$ and (b) $N_{tx}=32$, $N_{rx}=1$.}
\label{fig:Decoded_Constellations}
\end{figure*}

Consistency is critical for ultra-reliable communications, as reducing variance ensures more stable and predictable performance. Figs. \ref{fig:NN_vs_massive_MSE} and \ref{fig:NN_vs_massive_MSE_with_variance_bands} demonstrate that the NN decoder with precoding achieves lower normalized mean square error (NMSE) during training and validation compared to other methods. While higher-order antenna configurations show a higher error stabilization value for NMSE convergence, the 32-antenna setup exhibits significantly lower variance compared to the 4-antenna setup. This behavior can be attributed to differences in the distribution of the received QAM constellations for the $4\times1$ and $32\times1$ antenna arrangements.

More precisely, examining Figs. \ref{fig:Decoded_Constellation_4x1} and \ref{fig:Decoded_Constellation_32x1}, the $4\times1$ constellation appears less dispersed around the target symbols but features edge clusters that could result in higher BER during decoding. Conversely, the $32\times1$ configuration shows no edge clusters but exhibits greater dispersion overall. This suggests that NMSE might not be the most suitable metric for this evaluation step. Future work could explore the use of mutual information as an alternative evaluation metric.




\section{Conclusions}
\label{sec:conclusions}

This paper proposed an ultra-reliable MIMO communication scheme that integrates quasi-orthogonal space-time block coding (QOSTBC) with singular value decomposition (SVD)-based precoding to enhance system performance and robustness. By aligning transmissions with the dominant singular vectors of the channel, the proposed method mitigates the diversity loss inherent in QOSTBC, achieving performance comparable to traditional orthogonal STBC (OSTBC) while maintaining higher spectral efficiency. To address the computational complexity of QOSTBC decoding, we introduced a neural network-based decoder using phase-transmittance radial basis function (PT-RBF) architecture. This neural decoder jointly estimates and decodes the received signals, eliminating the need for separate channel estimation and reducing complexity. Simulations demonstrated that the neural network decoder closely approaches the performance of the optimal maximum likelihood (ML) decoder with perfect CSI, even in massive MIMO configurations with up to 32 transmitting antennas. Therefore, the combination of SVD-based precoding and PT-RBF neural network decoding offers a practical solution for ultra-reliable communication in next-generation networks. The proposed scheme enhances system reliability and spectral efficiency while ensuring consistent performance with low variability, essential for applications requiring high reliability. Future work will focus on optimizing the neural network architecture and training algorithms to improve performance and reduce computational requirements. Additionally, we plan to extend the analysis to higher-order modulations and investigate the impact of channel estimation errors on overall system performance.

\balance
\bibliographystyle{myIEEEtran.bst}
\bibliography{references.bib}

\end{document}